\documentclass[12pt]{article}
\usepackage{graphicx}
\usepackage[latin1]{inputenc}
\usepackage{epsfig}
\usepackage{amsmath}
\usepackage{amsfonts}
\usepackage{amssymb}
\usepackage{perpage} %the perpage package
\usepackage{slashed}
\usepackage{float}
\usepackage{color}
\usepackage{subfig}
\usepackage[table,xcdraw]{xcolor}
\usepackage{cite}
\MakePerPage{footnote}
\def\lsim{\:\raisebox{-0.5ex}{$\stackrel{\textstyle<}{\sim}$}\:}
\def\gsim{\:\raisebox{-0.5ex}{$\stackrel{\textstyle>}{\sim}$}\:}

\newcommand{\pz}{\partial_z}

\newcommand{\plmu}{\partial_{\mu}}

\newcommand{\dz}{D_z}

\newcommand{\dlmu}{D_{\mu}}

\newcommand{\BT}{\rule{0pt}{3ex} \rule[-2ex]{0pt}{0pt}}

\newcommand{\colmatt}[1]{\left(\begin{array}{cc}#1\end{array}\right)}

\newenvironment{institutions}[1][2em]
  {\begin{list}{}{\setlength\leftmargin{#1}\setlength\rightmargin{#1}}\item[]}
  {\end{list}}

\begin{document}

\begin{center}

	{	% TITLE GOES HERE 	
		\LARGE \bf 
		Tree-level Quartic for a Holographic Composite Higgs
	}

	\vskip .7cm
	
	\renewcommand*{\thefootnote}{\fnsymbol{footnote}}
	% Sets footnotes to a sequence of symbols, for e-mails

	{	% AUTHORS GO HERE
		\bf
		Csaba Cs\'aki$^{a}$, Michael Geller$^{b}$, and Ofri Telem$^{a}$ 
	}  

	\vspace{.2cm}

\begin{institutions}[2.25cm]
\footnotesize

    % UNIVERSITY IDENTIFICATIONS HERE

	$^{a}$ {\it Dept.\ of Physics, \textsc{lepp}, Cornell University, Ithaca, NY 14853}  
	\\
 	$^{b}$ {\it Dept.\ of Physics, University of Maryland, College Park, MD 20742} 
 
\end{institutions}	
\end{center}

%% Hacked a bit to fit where it should
\vspace{-1.5em}
\begin{center}{ \footnotesize\tt csaki@cornell.edu, mic.geller@gmail.com, t10ofrit@gmail.com} 
\end{center}

%%%%%%%%%%%%%%%%%%%%%
%%%  ABSTRACT    %%%%
%%%%%%%%%%%%%%%%%%%%%

\begin{abstract}
\noindent 
We present a new class of composite Higgs models where an adjustable tree-level Higgs quartic coupling allows for a significant reduction in the tuning of the Higgs potential. Our 5D warped space implementation is the first example of a holographic composite Higgs model with a tree-level quartic. It is inspired by a 6D model where the quartic originates from the ${\rm Tr} [A_5,A_6]^2$ term of the gauge field strength, the same model that led to the original little Higgs construction of Arkani-Hamed, Cohen, and Georgi. Beyond the reduction of the tuning and the standard composite Higgs signatures, the model predicts a doubling of the KK states with relatively small splittings as well as a Higgs sector with two doublets in the decoupling limit.  
\end{abstract}

\section{Introduction}

The origin of the Higgs potential and its stabilization is one of the key mysteries posed by the standard model (SM) of particle physics. An exciting possibility for explaining the dynamics behind electroweak symmetry breaking (EWSB) is that the Higgs boson itself is composite~\cite{Kaplan:1983fs,Georgi:1984af,Dugan:1984hq}, due to an additional strong interaction at scales about a decade or two above the weak scale. While this idea is intriguing, it does not work without additional structure: in order to reduce the scale of the Higgs mass well below the new strong coupling scale, one also needs to assume that the Higgs is a pseudo-Nambu-Goldstone boson (pNGB) of a global symmetry broken at a scale $f$, giving rise to pNGB composite Higgs models. There are two basic types of these: little Higgs (LH) models~\cite{ArkaniHamed:2001nc,LH} which were very popular in the early 2000s and (holographic) composite Higgs (CH) models~\cite{Agashe:2004rs,Contino:2006qr,SILH}, the simplest of which is the so called Minimal Composite Higgs Model (MCHM). For reviews see~\cite{Bellazzini:2014yua}. In both cases the essential ingredient for the 1-loop cancelation of the quadratic divergences is collective symmetry breaking~\cite{LH}, in which no single explicit breaking term breaks the global symmetry completely, and the divergences in the Higgs potential are softened. Most LH models contain a tree-level collective quartic (and a loop-induced finite or at most log divergent quadratic term), resulting in completely natural EWSB with no tuning. However, since the size of the quartic is determined by the same parameters as the quadratic, these models predict a heavy Higgs boson well above the observed $125$ GeV mass. Holographic composite Higgs models have a loop-induced quartic and therefore predict the correct size of the Higgs mass. This, however, comes at a cost of a $(v/f)^2$ tuning~\cite{Csaki:2008zd} in the Higgs potential. Additionally, the top partners in these models tend to be at least as heavy as $1.5-2\, \text{TeV}$ and thus not immediately discoverable at the LHC.

The tuning in holographic composite Higgs models could clearly be reduced if a tree-level but adjustable quartic were present.\footnote{For an alternative recent approach towards reducing the tuning in the Higgs potential for CH models see~\cite{Csaki:2017cep}.} This realization has inspired us to revisit the original little Higgs model~\cite{ArkaniHamed:2001nc}, formulated as a 6D gauge theory where two Higgses correspond to two Wilson loops going around the fifth and sixth dimensions and the collective quartic arises from the field strength term ${\rm Tr} [A_5,A_6]^2$.

The aim of this paper is to implement the ideas of~\cite{ArkaniHamed:2001nc} within the holographic approach where the extra dimension is warped. For this purpose, we construct a 6D model on an $AdS_5\times S_1$ background where the quartic is generated similarly to~\cite{ArkaniHamed:2001nc} and the Higgs can be interpreted as a composite pNGB. We discuss the essential aspects of the 6D model and then quickly zoom in on a simple and transparent formulation in terms of a warped 5D model, where only the sixth dimension has been deconstructed. The resulting Higgs sector is a CP-conserving two Higgs doublet model (2HDM) in the decoupling limit with a tree-level, MSSM-like quartic. As we will see, this quartic can be adjusted to fit the observed value without extra tuning.

In its warped 5D version it provides the first example of a composite Higgs model with a tree-level Higgs quartic coupling in which the only source of tuning is related to the reduction of the Higgs mass parameter. Moreover, the top partners in this model can be light and discoverable at the LHC. It turns out to be a relatively simple model which captures almost all the essential elements of the 6D theory (as well as the original model of~\cite{ArkaniHamed:2001nc}).

The paper is organized as follows: Sec.~\ref{sec:tuning} contains an explanation of the reduction of the tuning in the Higgs potential due to the presence of the adjustable Higgs quartic. In Sec.~\ref{sec:6D} we present the essential ingredients of the 6D theory and the structure of the zero modes. Sec.~\ref{sec:5D} contains the warped 5D model, which is the main new result of this paper. We provide the matter content along with the structure of the Higgs potential and a mechanism for lifting the flat direction in the tree-level potential in Sec.~\ref{sec:SMmatter}. The matching onto generic 2HDM models is contained in Sec.~\ref{sec:2HDM}, and the basic elements of the expected phenomenology in Sec.~\ref{sec:pheno}. We conclude in Sec.~\ref{sec:Conclusions}.

\section{Motivations for a Quartic from 6D \label{sec:tuning}}

The first implementation of the little Higgs idea~\cite{ArkaniHamed:2001nc} was based on a  deconstructed~\cite{ArkaniHamed:2001ca} 6D gauge theory. The aim was to construct a composite Higgs model where a large  tree-level quartic could result in a fully natural electroweak symmetry breaking (EWSB) Higgs potential. The extra dimensional components $A_5,A_6$ of the gauge field can have the right quantum numbers to be identified with the Higgs. Compactification of the extra dimension can provide physical irreducible Wilson lines in the extra dimension which have all the properties of a pNGB in 4D (see also \cite{Gregoire:2002ra}). The quartic arises from the field strength term:
\begin{equation}
 {\rm Tr} [A_5,A_6]^2  \in F_{56} F_{56}.
\end{equation}
In the deconstructed version, this corresponds to a plaquette operator. 

Before explaining the details of the full 6D construction (as well as the simple warped 5D version), we would like to explain how the presence of such a tree-level quartic could help alleviate the tuning in composite Higgs models\footnote{For a detailed analysis of the tuning in CH models, see \cite{Panico:2012uw}.}. The Higgs potential in CH models with a loop-induced quartic is parametrized as 
\begin{equation}
V(h) = \frac{3 g_t^2 M^2_{\Psi}}{16 \pi^2} \left(-a h^2+\frac{b}{2}\frac{h^4}{f^2}\right)
\end{equation}
where $g_t $ is the SM coupling, $M_{\Psi}=g_\Psi f$ is the top partner mass, and $a$ and $b$ are (at most) ${\cal O}(1)$ numbers. The coefficients $a$ and $b$ can be smaller than 1 (at the price of tuning various terms against each other) but can not be bigger than ${\cal O}(1)$. The tuning is then quantified by 
\begin{equation}
\Delta = \frac{1}{ab}\ .
\end{equation}
The origin of the $v^2/f^2$ tuning is easy to see: since both the quadratic and quartic terms are loop-induced by the same dynamics, the minimum of the potential is when $\frac{v^2}{f^2} = \frac{a}{b}$, which for $b \sim 1$ gives the ``irreducible" tuning of composite Higgs models $\Delta = \frac{1}{ab} = \frac{f^2}{v^2} \gsim 9$. The lower bound on this tuning follows from electroweak precision and Higgs coupling constraints, which imply that that $\frac{f}{v} \gsim 3$. A more detailed analysis of the tuning yields 
\begin{equation}\label{eq:indr}
\Delta \simeq 8 \, y_t \left( \frac{g_\Psi}{1.8}\right)^2 \left( \frac{f/v}{3} \right)^2 \gsim 8 \, y_t\ ,
\end{equation}
since $g_\Psi>1.8$ is required to get a large enough loop-induced quartic.

An additional (adjustable) tree-level quartic significantly changes the picture, reducing the previously ``irreducible" $\frac{f}{v}$ tuning. The reason is that the coupling $g_\Psi$ setting overall magnitude of the loop-induced Higgs potential can be taken smaller, while the adjustable contribution ensures that $\lambda=0.13$. In this case the dominant bound on the tuning is no longer the indirect bound Eq.~(\ref{eq:indr}) but rather the direct bound on the top partner mass from direct searches at the LHC,
\begin{equation}
\Delta \simeq 5.5\,y_t\, \left( \frac{M_{\Psi}}{1100\ {\rm GeV}} \right)^2\gsim 5.5\, y_t\, .\label{eq:tuning}
\end{equation}
This is parametrically weaker than the bound in the conventional CH. Moreover, Eq.~\ref{eq:tuning} implies that top partners should lie just around the corner, well within the reach of the LHC. This is contrary to the standard CH case, where light top partners are disfavored by Higgs and EW data \cite{Panico:2012uw}.

The reduction of the bounds on tuning in twin Higgs type models~\cite{Chacko:2005pe,TH2,Craig,Geller:2014kta,Low:2015nqa,Barbieri:2015lqa,newTH,Craig:2015pha} is even more impressive. One can show that twin Higgs models with an additional source of quartic will remain natural (no tuning required) even after the end of the high luminosity 13 TeV run of the LHC. Similar ideas have been discussed in \cite{Harnik:2016koz,Katz:2016wtw}.

\section{The 6D Composite Higgs Model\label{sec:6D}}

In the following section we present a warped model with a collective tree-level quartic for the pNGB Higgs. This model can either be written in six dimensions or deconstructed into lower dimensional models. As we will see in the next section, the simplest and most useful representation of this model is as a warped 5D model with two sites in the bulk representing the sixth dimension. To motivate this construction, we first review the essential features of the full 6D theory. 
 
\begin{figure}
\vspace*{-0.5cm}
\begin{center}
\includegraphics[scale=0.25]{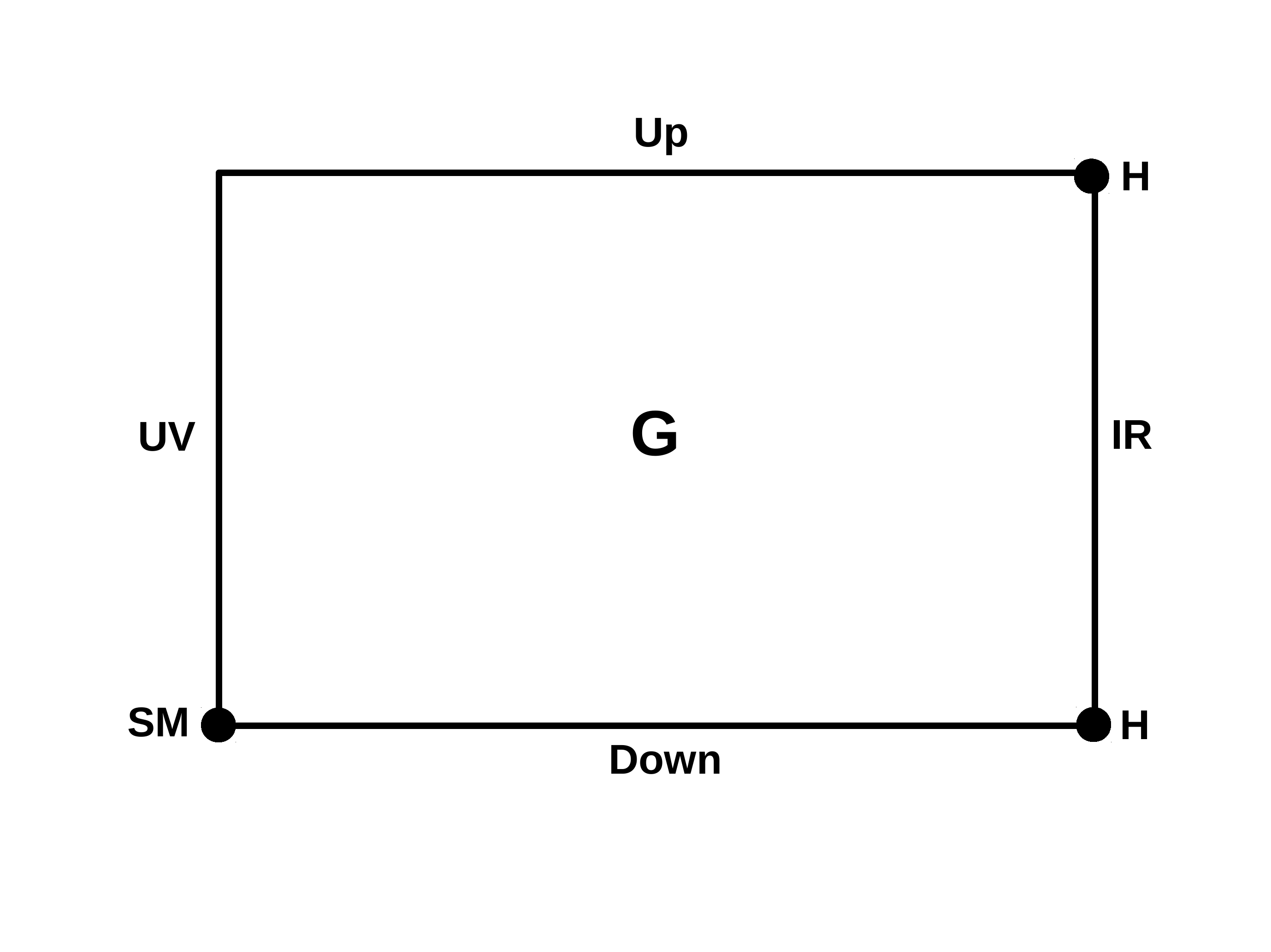}
\end{center}
\vspace*{-1cm}
\caption{ A sketch of the layout of the 6D model. The rectangle represents the two extra dimensions, the horizontal corresponding to the warped extra dimension, the vertical to the extra flat segment of the 6th dimension. The 3-branes in the 3 corners represent the symmetry breaking pattern at those locations, necessary to obtain the appropriate pattern of Higgs fields and couplings. \label{fig:6Dfi}}
\end{figure}

The 6D model is defined on an interval of $AdS_5\times S_1/Z_2\times Z_2$ with metric 
\begin{equation}
ds^2 = \left(\frac{R}{z}\right)^2 (dx^2 -dz^2) -dy^2
\end{equation}
where the $R<z<R^\prime$ coordinate parametrizes the warped direction of $AdS_5$ and $0<y<R_y$ parametrizes the $S_1$ direction. The geometry along with the boundary conditions is illustrated in Fig.~\ref{fig:6Dfi}. At $z=R$ and $z=R'$ there are UV and IR 4-branes, and at $y=0$ and $y=R_y$ there are other 4-branes, denoted Down and Up. Furthermore, there are 3-branes at the four `corners' UV-Down, UV-Up, IR-Up, and IR-Down. 

The bulk gauge symmetry of the model is $G$ (usually chosen to be $SO(5)$ to incorporate custodial symmetry), broken to an $H$ subgroup ($SO(4)$ in the simplest model) on the IR-Up, and IR-Down 3+1 branes. This setting is a direct extension of the standard CH construction of a bulk $G$ in $AdS_5$ broken to $H$ on the IR brane. The two IR symmetry breaking points in our case correspond to two sets of $G/H$ pNGBs instead of just a single one in the standard CH.
Additionally, the bulk symmetry on the UV-Down corner to the SM $SU(2)\times U(1)$. This is analogous to the UV breaking in the standard CH, which ensures that only the SM gauge bosons remain light.

In a similar way to 5D gauge-Higgs unification, the Higgses in our setting arise as the two Wilson lines connecting the UV-down corner where the G symmetry is broken with the two IR symmetry breaking points. To calculate the Higgs potential, we start from the 6D Maxwell action
\begin{equation}
\mathcal{S} =  -\frac{1}{4g_6^2}\int d^4 x dz dy \, \sqrt{g} F^a_{MN}F^{a;MN}\ .
\end{equation} 
% \int dy  \int dz  \frac{R^3}{z^3} \left(F^a_{\mu \nu} F^{a,\mu \nu} +\frac{z^2}{R^2} F^a_{\mu z} F^{a,\mu z}  + F^a_{\mu y} F^{a,\mu y}  + F^a_{z y} F^{a,z y} \right),
%\end{equation}
Choosing a bulk Lorentz gauge ($\partial^M A^a_{M}=0$), we find the bulk equations of motion for $A^a_{z,y}$ simplify to:
\begin{equation}\label{eq:6Deom}
\pz \left(\frac{R^3}{z^3}  F^a_{z y}\right)=0~,~\partial_y\left(\frac{R^3}{z^3}  F^a_{z y}\right) =0 \, .
\end{equation}
By integrating the bulk action by parts, we can also obtain the boundary conditions:
\begin{equation}
F^a_{zy}~|_{UV,~IR,~UP,~DN}=0~,~F^a_{\mu z}~|_{UV,~IR}=0~,~F^a_{\mu y}~|_{UP,~DN}=0 \ .
\end{equation}
We would like to emphasize that these BC are valid at generic points on the 4-branes, but not on the 3-branes at the UV-Down, IR-Up, and IR-Down corners in the corners, where the BC are modified for the broken generators. In the Lorentz gauge the generic BC can be rewritten as:
\begin{equation}
F^a_{zy}~|_{UV,~IR,~UP,~DN}=0~,~A^a_{z}~|_{UV,~IR}=0~,~A^a_{y}~|_{UP,~DN}=0.
\end{equation}
Together with the bulk EOMs these BCs can only be satisfied if $F^a_{z y}=0$ throughout the entire bulk. The vanishing of $F^a_{z y}\equiv \left(\partial_yA^{a}_z - \pz A^a_y\right)$ allows us to define a bulk potential $F^a$ so that $A^a_z = \pz F^a~,~ A^a_y=\partial_yF^a$. The potential $F^a$ satisfies the warped version of a 2D Laplace equation, and will be the main object of interest for us. This potential also has a distinct physical meaning: after fixing an integration constant, this potential is exactly the log of the Wilson line from the UV-Down 3-brane to any other point in the bulk. 

\begin{figure}
\centering
\subfloat[]{\centering \includegraphics[height=4cm]{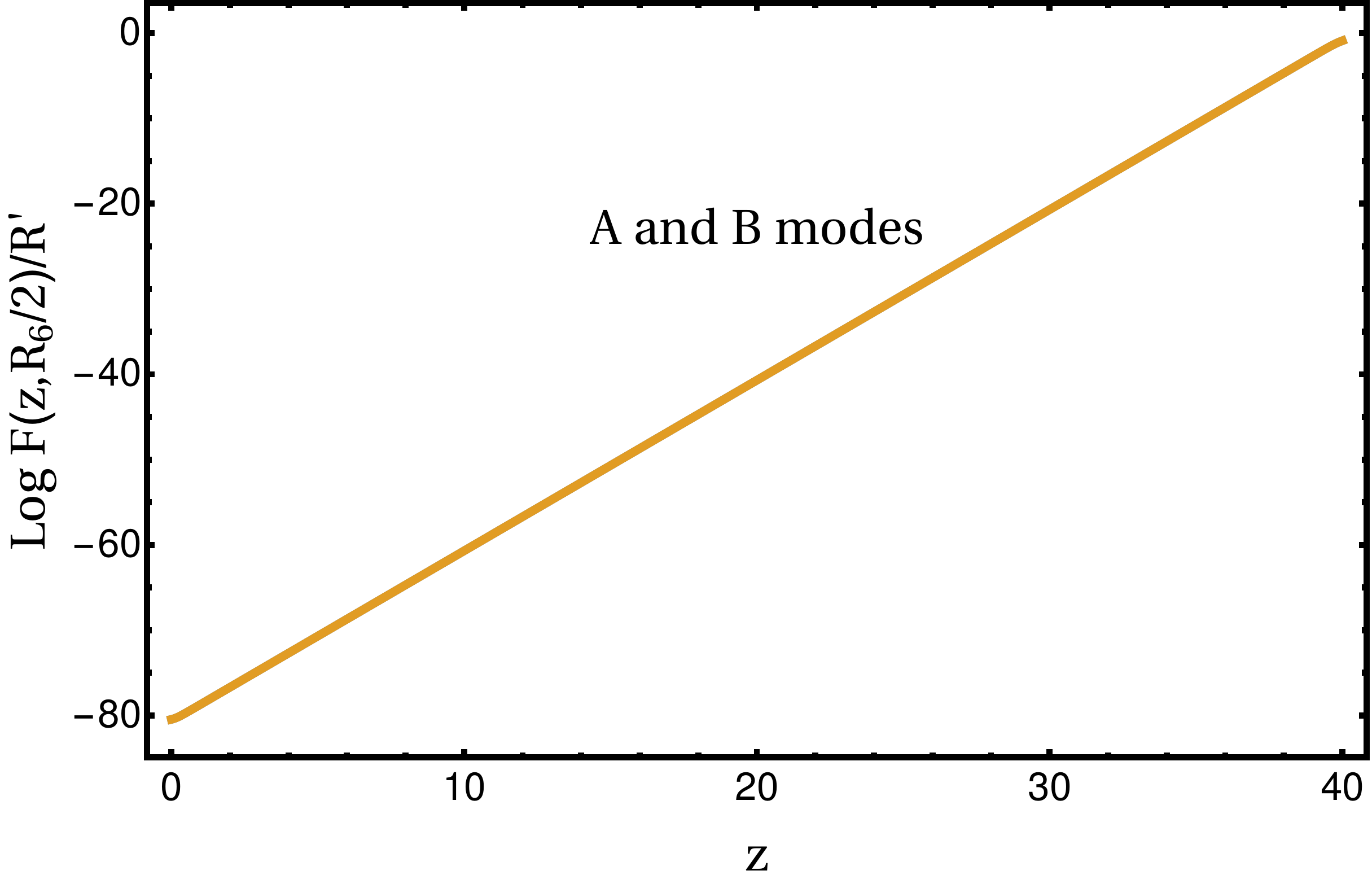}}
\hspace*{0.2cm}
\subfloat[]{\centering \includegraphics[height=4cm]{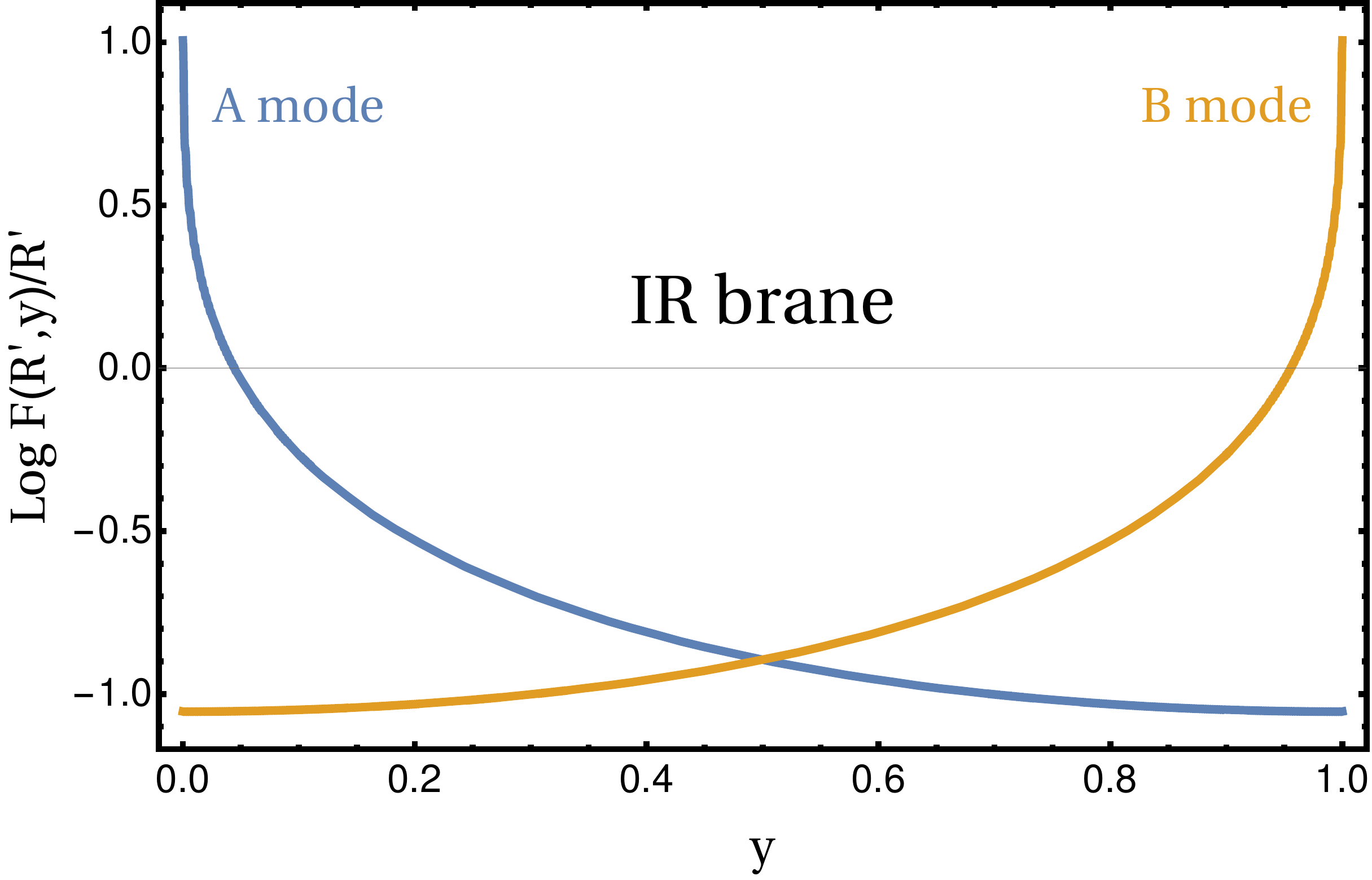}}
\caption{Slices of the 6D zero modes. (a): z-slice at $y=R_6/2$ as a function of $z$, illustrating that both are IR localized. (b) $y$-slice of the IR 4-brane $z=R'$.  Here the modes differ: mode A is localized close to the IR-Down corner, while mode B is localized close to the IR-Up corner.
\label{fig:mode}}
\end{figure}

The problem of finding zero modes for $A_z$ and $A_y$ is thus reduced to solving the warped Laplace equation on a rectangle with Dirichlet boundary conditions everywhere but the three gauge symmetry breaking points on the UV-Down, IR-Up, and IR-Down corners. We have obtained the full solution to this problem via Fourier transforms.
The solution is a linear combination of two zero modes $A,B=1,2$: 
\begin{equation}
F^i(z,y,x)=F_{A}(z,y )~h^i_{A}(x)~+~F_{B}(z,y)~h^i_{B}(x),
\end{equation}
where  $F_{A}$ are the extra dimensional wave functions and $h^i_{A}$ are two 4D doublet modes. Both of these are IR localized, $F_{A}$ is localized in the IR-Up corner and $F_{B}$ is localized in the IR-Down corner. The index $i$ above stands for the $SU(2)_{L}$ doublet embedded in the adjoint of $SO(5)$ as usual, and will be suppressed in the following. We present slices of the zero modes obtained from solving the 2D Laplace equation in Fig.~\ref{fig:mode}. We can see that indeed both modes are IR localized, and peak at two different corners on the IR brane. 

\section{A 5D Model Holographic Composite Higgs Model with a Tree-level Quartic\label{sec:5D}}

\begin{figure}
\begin{center}
\includegraphics[scale=0.3]{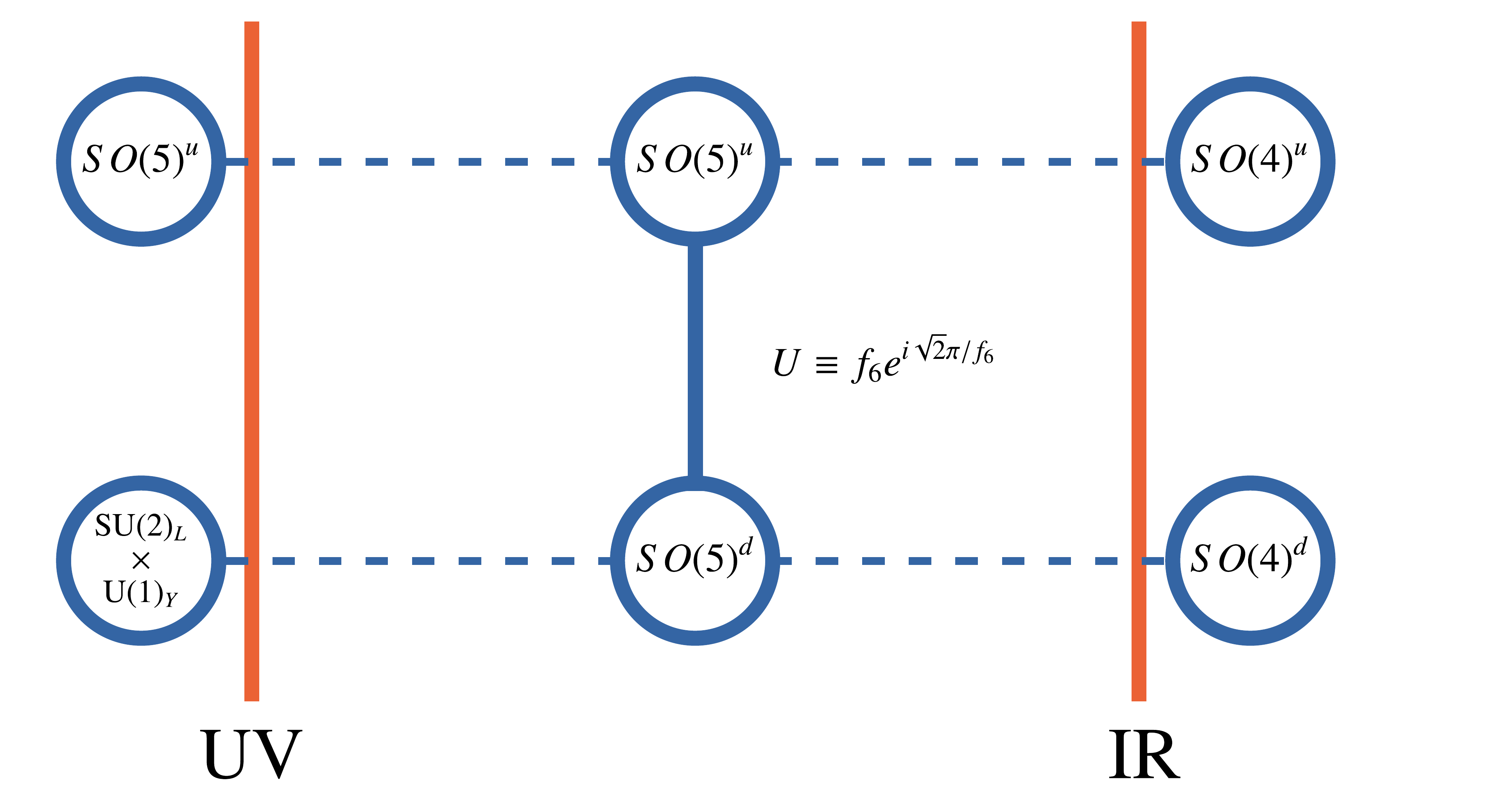}
\end{center}
\caption{A sketch of the main elements of our deconstructed 5d model. The two sites in the bulk represent the $SO(5)^u\times SO(5)^d$, broken on the IR into $SO(4)^u\times SO(4)^d$ and on the UV into $SO(5)^u\times {\left[SU(2)_L\times U(1)_Y\right]}^d_{SM}$. The bulk link corresponds to the breaking of $SO(5)^u\times SO(5)^d$ into the diagonal $SO(5)_V$ with a constant VEV in the bulk. }\label{two_site_cartoon}
\end{figure}

In this section we present the key result of this paper: a warped 5D composite Higgs model with a non-vanishing tree-level quartic for the Higgses. It is based on the 6D model outlined in the previous section, by deconstructing the flat sixth dimension as a two-site model, while the warped fifth dimension is kept unchanged. For previous attempts at warped UV completions of LH models see~\cite{Thaler:2005en,Csaki:2008se}. The background is the standard AdS$_5$ metric in the 5D bulk 
\begin{equation}
ds^2=\frac{R^2}{z^2}\left(dx^2-dz^2\right)
\end{equation}
and the bulk Lagrangian looks like a two-site model: the bulk gauge symmetry is $SO(5)^u\times SO(5)^d$, where $i=u,d$ are the two sites mimicking the effect of the 6th dimension. The bulk symmetry is broken on the IR brane to $SO(4)^u\times SO(4)^d$ and on the UV brane in $SO(5)^u\times {\left[SU(2)_L\times U(1)_Y\right]}^d_{SM}$. In addition, the $SO(5)^u\times SO(5)^d$ symmetry is broken in the bulk to the diagonal $SO(5)_V$, with the original $SO(5)^u\times SO(5)^d$ realized nonlinearly via a bulk link field $U=e^{i\frac{\sqrt{2}}{f_6}\pi^a T^a}$. The $\pi^a$'s play the role of the 6th component of the 6D $SO(5)$ gauge fields $A_y$. The decay constant $f_6$ is taken to be a constant along the 5th dimension, and is roughly of the order of the inverse AdS curvature $1/R$. This model is illustrated in Fig.~\ref{two_site_cartoon}.

Next we will explicitly show that this model contains two scalar zero modes with IR localized profiles. The bulk action for this model includes the gauge kinetic terms and the covariant derivative of the link field:

\begin{equation}\label{eq:zlag}
S=\int d^4x dz \sqrt{g} \left( -\frac{1}{4}F_{MN}F^{MN}  + \frac{f_6^2}{4} \left(D^M U\right)^\dagger D_M U  \right),
\end{equation}
where the covariant derivatives are
\begin{equation}\label{eq:dz}
 \dlmu U = \left(\frac{\sqrt{2}}{f_6} \partial_\mu \pi - g_5A^{(A)}_\mu \right)  +... , \ \ 
  \dz U = -\left(\frac{\sqrt{2}}{f_6}\pz \pi - g_5 A^{(A)}_z - i\frac{\sqrt{2} g_5}{f_6}  \left[A^{u}_z,\pi\right] \right)  +... 
\end{equation}
and $A^{(A)}_{\mu,z} \equiv A^u_{\mu,z} - A^d_{\mu,z}~,~A^{(V)}_{\mu,z} \equiv A^u_{\mu,z} + A^d_{\mu,z}$, while the ellipses stand for higher order commutator terms and terms negligible in the limit $R \ll R'$.
The scalar zero modes live in a linear combination of  $\pi$ and $A^{u,d}_z$. We will use the Lorentz gauge $\partial_\mu A^\mu = 0$, with the additional gauge fixing $A^{(A)}_0=0$. The vector combination $A^V_{z}$ corresponds to an unbroken bulk symmetry, and so the part of the zero mode contained in the  corresponding fifth component must  be of the usual form $A^V_{z}\sim z$. Substituting the $\pi$ variation in the axial part of the $A_0$ variation, we obtain the equation for the zero mode as well as the relation between $A_z$ and $\pi$ components of the zero mode:
\begin{eqnarray}
\frac{z^3}{R^3}\pz\left[\frac{R}{z}\, \pz \pi\right]-g^2_5f^2_6 \, \pi&=&0\nonumber\\
\pz \pi ~~~- ~\frac{g_5f_6}{\sqrt{2}}A^{Ax}_z&=& 0\, .
\end{eqnarray}
We choose the UV boundary condition to be $A^u_z |_{UV}= 0$.  We do not need to  specify an IR boundary condition for the $A_z$s: the reason is that in the gauge we chose there is no boundary localized term involving $A_z$, thus one does not need to consider the boundary variation of these fields. This is in accordance with our expectations from the fact that the $A_z$ equations of motion are effectively first order, hence one BC should be sufficient.  The space of zero mode solutions is two dimensional, spanned by modes A and B satisfying:
\begin{eqnarray}
\text{A mode} &:& \ \ A^d_z |_{IR}= 0\nonumber\\
\text{B mode} &:& \ \ A^u_z |_{IR}= 0  \, .
\end{eqnarray}

The two solutions to the EOM are:
\begin{eqnarray}
{\pi(z,x)}_{A/B} &=&   \, \frac{g_5f_6}{2\sqrt{2}}  \left[\pm \frac{1}{1+\alpha}\, {\left(\frac{R}{R'}\right)}^{\alpha}\,{\left(\frac{z}{R}\right)}^{1+\alpha}+ \frac{1}{1-\alpha}\,\left(\frac{R}{R'}\right)\,{\left(\frac{z}{R}\right)}^{1-\alpha}\right] h_{A/B}(x)\nonumber \\
{A^{u}_z(z,x)}_{A/B}  &=& \frac{1}{2R} \left\{\frac{z}{R'}  + \left[\pm {\left(\frac{R}{R'}\right)}^{\alpha} \, {\left(\frac{z}{R}\right)}^{\alpha}+\left(\frac{R}{R'}\right) \, {\left(\frac{z}{R}\right)}^{-\alpha}\right]\right\}h_{A/B}(x)\nonumber \\
{A^{d}_z(z,x)}_{A/B}  &=& \frac{1}{2R} \left\{\frac{z}{R'}  - \left[\pm {\left(\frac{R}{R'}\right)}^{\alpha} \, {\left(\frac{z}{R}\right)}^{\alpha}+\left(\frac{R}{R'}\right) \, {\left(\frac{z}{R}\right)}^{-\alpha}\right]\right\}h_{A/B}(x)\, ,
\end{eqnarray}
where $\alpha=\sqrt{1+g^2_5f_6^2R^2}$ and $h_A(x)$ and $h_B(x)$ are the two canonically normalized 4D modes. The $\pm$ mean a "$+$" for the A mode and a "$-$" for the B mode. 

In the $g_5f_6R=g_* f_6 R^{3/2}\equiv {\theta}_6\ll 1$ limit, which is necessary to obtain a quartic smaller than $\mathcal{O}(1)$ (see Sec.~\ref{sec:5Dquartic}), the profiles far from the UV brane ($z\gg R$) are:
\begin{eqnarray}\label{eq:approx}
\text{A Mode}&:& A^u_z(z)= \frac{1}{R}\frac{z}{R'}\, h_{A}(x)~,~A^d_z(z) ={\cal O}({\theta}^2_6)\, h_{A}(x)~,~\pi(z)= {\cal O}({\theta}_6)\, h_{A}(x)\nonumber\\
\text{B Mode}&:& A^u_z(z) = {\cal O}({\theta}^2_6 )\, h_{B}(x)~,~A^d_z(z) =\frac{1}{R}\frac{z}{R'}\, h_{B}(x)~,~\pi(z)= {\cal O}({\theta}_6)\, h_{B}(x) 
\end{eqnarray}
This can be understood as the limit of no bulk breaking, in which the two modes simply correspond to the pNGBs of the two separate SO(5)/SO(4) cosets. It is then sensible to define the pNGB matrices:
\begin{equation}
\Sigma^u = \exp\left(i g_6 \int_R^{R'}A^u_z dz \right)\equiv e^{i h_u/f}~,~\Sigma^d = \exp\left(i g_6\int_R^{R'}A^d_z dz \right)\equiv e^{i h_d/f}\label{eq:pNGBs}
\end{equation}
These matrices restore the two $SO(5)$ symmetries by rotating the IR vacua that break $SO(5)$ into $SO(4)$. In the limit $ {\theta}_6\ll 1$:
\begin{equation}\label{eq:ident}
h_u \approx h_A~,~h_d \approx h_B\, .
\end{equation}
Since we need to work in this limit for any realistic model building, from now on we identify $h_{A,B}$ with $h_{u,d}$. 
\subsection{The tree-level quartic \label{sec:5Dquartic}} 
We can now evaluate  the quartic scalar coupling by substituting the zero modes in the commutator term of Eq.~(\ref{eq:dz}):
\begin{eqnarray}
\mathcal{L}_{\lambda} ~ &=& \lambda \, \text{Tr} \left[h_A,h_{B}\right]^2\approx  \lambda \, \text{Tr} \left[h_u,h_d\right]^2
\end{eqnarray}
with
\begin{eqnarray}
\lambda &=& g^2_5 \int dz \, {\left(\frac{R}{z}\right)}^3  \, {\left[\pi_A \,  A^V_{z,B} - \pi_{B} \, A^V_{z,A}\right]}^2 \, .
\end{eqnarray}
Plugging in the profiles and working in the limit $R'/R \rightarrow \infty$, we get:
\begin{eqnarray}
\lambda ~ &=& g^2_5\frac{g^2_5f^2_6R}{4{\left(1+\alpha\right)}^3}\sim \frac{1}{4}g^2_*\,\theta^2_6 \, . \label{Eq:5Dquartic}
\end{eqnarray}
where the last result is in the limit that the scale of the bulk breaking is small, i.e. $\theta_6 \ll 1$. This choice is necessary to get a quartic which is smaller than ${\cal O}(1)$, and as promised, also justifies the identification Eq.~(\ref{eq:ident}).

The quartic has the following structure:
\begin{equation}\label{eq:qrt}
\lambda \left[h_u,h_d\right]^2 = \lambda \left(h_1^\dagger h_1 - h_2^\dagger h_2\right)^2 +\lambda \text{ tr}\left(h_1h_1^\dagger - h_2h_2^\dagger\right)^2 \, ,
\end{equation}
where we have adopted the definitions of \cite{ArkaniHamed:2001nc}:
\begin{equation}\label{eq:ark}
h_1 = \frac{1}{\sqrt{2}}\left(h_u + i h_d \right)~,~h_2 = \frac{1}{\sqrt{2}}\left(h_u- i h_d \right) \, .
\end{equation}

This quartic has a flat direction for $h_u$ or $h_d$, which corresponds to $h_1= \pm h_2$. In Sec.~\ref{sec:lift} we will discuss how these flat directions are lifted radiatively by the SM top and its partners.

\subsection{The 4D interpretation}

The warped 5D picture makes it easier to identify the 4D dual to our model based on the rules of the AdS/CFT correspondence~\cite{AdSCFT}. As in the standard CH, this dual is a strongly interacting theory which is close to conformal in the UV but confines in the IR. Above the confinement scale there is a global symmetry $G\times G$ (the bulk symmetry in the 5D picture), and the IR condensate breaks it to $H\times H$ with two sets of $G/H$ pNGBs. 

In addition, the CFT is coupled to a UV source which explicitly breaks the $G\times G$ symmetry to the diagonal $G$. In the CFT, a composite operator $\mathcal{O}_{G\times G}$ that transforms under the global $G\times G$ couples to that source and its dimension is related to $f_6$, the scale of the breaking in the bulk. We stress that the bulk scale $f_6$ does not directly correspond to a 4D scale but rather to a 4D scaling dimension, just like bulk fermion masses in RS~\cite{John}.

Due to the explicit breaking to the diagonal $G$, the pNGBs get a tree-level potential. Remarkably, this tree-level potential consists only of a commutator-squared term, and in particular does not include tree-level masses for the pNGBs. This is a demonstration of the 'collective' nature of our potential. 

\section{The SM field content\label{sec:SMmatter}}

So far we have only considered the scalar modes and their tree-level quartic (while also introducing the SM gauge fields). In this picture, the 4D gauge fields correspond to the zero modes of the surviving diagonal subgroup of the bulk gauge symmetry: 
\begin{equation}
A_\mu^{SM}= A_\mu^{u}+A_\mu^{d}.
\end{equation}
These are the $SU(3)\times SU(2)\times U(1)_Y$ gauge fields, which have Neumann BC on both branes and no bulk mass. 

The scalar sector consists of two pNGB Higgs scalars which in the $\theta_6 \ll 1$ limit are the $h_{u,d}$ of $SO(5)^{u,d}/SO(4)^{u,d}$. However, the $h_{u,d}$ basis is not the physical basis for our Higgses, as we will see from the radiatively generated contribution. Instead, we work in the $h_{1,2}$ basis of Eq.~(\ref{eq:qrt}), with the quartic given by Eq.~(\ref{eq:ark}).

Next we introduce the SM fermion matter content into this theory. Presently we do not specify the light fermion representations, as there are many different possibilities. The exact choice will definitely be relevant for flavor physics, but its effect on the Higgs potential is negligible. In the following we will only focus on the top sector and its contribution to the Higgs potential. 
To be consistent with the LHC results on Higgs couplings, we require one Higgs to be SM-like, while the other one should be heavy. We achieve this by arranging for the top to couple only to $h_1$ so that the top loop gives a negative mass term for $h_1$, while $h_2$ is lifted by the other members of the top's $SO(5)$ multiplet.

\subsection{The top sector}

The simplest choice for the top sector is to embed $t_L$ in a multiplet $T_L$ transforming as a $(\mathbf{5},\mathbf{1})+(\mathbf{1},\mathbf{5})$ of $SO(5)\times SO(5)$, while $t_R$ is a singlet localized on the IR brane. This is summarized in Tab.~\ref{tab:5Dreps}. In terms of two component spinors, the bulk fermion $T_L$ can be written as
\begin{eqnarray} 
T_L=\colmatt{\chi^u_L & \chi^d_L\\ \bar{\psi}^u_L & \bar{\psi}^d_L}\, ,
\end{eqnarray}
where $\chi^{u,d},\psi^{u,d}$ are 2-component Weyl spinors. We impose a $Z_4$ symmetry on the fermion sector in the bulk and on the $U$ field:
\begin{eqnarray} 
Z_4 :&&~~T^u_L\rightarrow  -T^d_L~~~~~~T^d_L\rightarrow  T^u_L\nonumber\\
&&~~~~~~~~U \rightarrow  U^\dagger \left(\pi \rightarrow -\pi\right)~. \label{eq:Z4}
\end{eqnarray}
As we will see, this $Z_4$ symmetry ensures that the top couples only to $h_1$, which is important for a realistic Higgs potential.
%%
%%\begin{eqnarray} 
%%T_L=\colmatt{\chi^u_L & \chi^d_L\\ \bar{\psi}^u_L & \bar{\psi}^d_L}~,~T^u_R=\colvec{\chi^u_R \\ \bar{\psi}^u_R }~,~T^d_R=\colvec{\chi^d_R \\ \bar{\psi}^d_R }
%%\end{eqnarray}
The 5D Lagrangian for $T_L$ can be written as
\begin{eqnarray} 
\mathcal{L}_{T_L}&=&{\left[i\bar{\chi}_L \bar{\sigma}^{\mu} \dlmu \chi_L+i \psi_L\sigma^{\mu} \dlmu \bar{\psi}_L + \frac{1}{2}\left(\psi_L \overleftrightarrow{\dz} \chi_L - \bar{\chi}_L \overleftrightarrow{\dz} \bar{\psi}_L\right)+\frac{c_L}{z}\left(\psi_L \chi_L+\bar{\chi}_L\bar{\psi}_L\right)\right]}^{u,d} + \nonumber\\
& & \frac{y_6 \theta_6 }{2z} \left(i\psi^d_L \, U^\dagger \,\chi^u_L-i\psi^u_L\,U\chi^d_L+i\bar{\chi}^d_L \, U^\dagger \,\bar{\psi}^u_L-i\bar{\chi}^u_L\,U\,\bar{\psi}^d_L \right)\, ,
\end{eqnarray}
where $y_6$ is a ${\cal O}(1)$ coupling of the link field. The first line is just the standard 5D warped fermion Lagrangian, while the second line includes the fermion couplings to the link field which stand for their coupling to $A_6$ in the full 6D Lagrangian. The bulk masses for both multiplets are assumed to be equal. This arises naturally in the 6D picture and can be thought of as a $u \leftrightarrow d$ $Z_2$ symmetry that is preserved in the bulk and on the IR brane (while broken in the UV). 

\begin{table}[h!]
\centering
\label{5D reps}
\begin{tabular}{|c|c|c|}
\hline
\rowcolor[HTML]{9B9B9B} 
 \BT           \textbf{Field} &             $\mathbf{SO(5)^u}$ &        $\mathbf{SO(5)^d}$ \\ \hline
 \BT       ${\left(\chi_L,\bar{\psi}_L\right)}^u$         &                $\square$          &            $\mathbf{1}$                       \\ \hline
 \BT       ${\left(\chi_L,\bar{\psi}_L\right)}^d$         &                 $\mathbf{1}$         &           $\square$     \\    \hline
  \BT       $t_R $         &                $\mathbf{1}$          &            $\mathbf{1}$                       \\ \hline
 \BT       U        &         $\square$                 &            $\bar{\square}$                        \\ \hline
\end{tabular}
\caption{Representations of the 5D Model Top Sector}\label{tab:5Dreps}
\end{table}

The $A_z$ VEV is set to zero by a gauge rotation \cite{Falkowski:2006vi}, with the $A_z$ Wilson line entering in the IR boundary condition. Rotating to the bulk mass basis, we have
\begin{eqnarray} 
\mathcal{L}_{T_L}&=&  {\left[i\bar{\chi}_L \bar{\sigma}^{\mu} \plmu \chi_L+i \psi_L\sigma^{\mu} \plmu \bar{\psi}_L + \frac{1}{2}\left(\psi_L \overleftrightarrow{\pz} \chi_L \right)+\frac{c_L\pm y_6 \theta_6}{z}\psi_L \chi_L\right]}^{0,1},
\end{eqnarray}
where $\psi^{0,1}_L=\frac{1}{\sqrt{2}}\left(\psi^u_L \pm i \psi^d_L\right),\chi^{0,1}_L=\frac{1}{\sqrt{2}}\left(\chi^u_L \pm i \chi^d_L\right)$. 

We see that the full $T_L$ bulk multiplet splits into two bulk fermions with masses $c_L\pm  y_6\theta_6$.
On the IR brane we will always take the same BC for the up and down fermions, while on the UV the BC may be different. To understand how this works, take for example the following BC for a bulk fermion:
\begin{equation}
\tilde{\chi}^u_L:(-,+) ~,~\tilde{\chi}^d_L:(+,+)\, ,
\end{equation} 
where the tilde serves to remind us that the boundary conditions are for the Wilson line rotated fields
\begin{equation}
\tilde{\chi}^u_L=\chi^u_L\, \Sigma^u~,~\tilde{\chi}^d_L=\chi^d_L\, \Sigma^d\, ,
\end{equation} 
and $\Sigma^{u,d}$ are the usual pNGB matrices given by Eq.~(\ref{eq:pNGBs}).
In terms of the two combinations in the bulk mass basis $\tilde{\psi}^{0,1}_l$ this corresponds to: 
\begin{eqnarray}
\tilde{\psi}^0_L(R) =  \tilde{\psi}^1_L(R)~&,&~\tilde{\chi}^0_L(R) =  -\tilde{\chi}^1_L(R)\nonumber\\ 
\tilde{\psi}^0_L(R') =~~~~0 ~~~~~~&,&\, \tilde{\psi}^1_L(R')=~~~0\,~~~~~~~~ .
\end{eqnarray}
These boundary conditions result in a single zero mode that lives both in the $0$ and $1$ parts of the multiplet:
\begin{equation}\label{eq:bulkk}
\chi^{0,1}_L(z,x) \, \sim\, {\left(\frac{z}{R}\right)}^{c_L \pm y_6\theta_6} t_L\left(x\right)\, .
\end{equation}
Note that the combination $\chi^0$ with the higher bulk mass rises faster towards the IR, and so the dominant contribution to the Higgs potential comes from that combination. In the following calculation of the Higgs potential we will neglect the $\chi^1$ contribution, and comment on it in the end of section~\ref{sec:2HDM}.

On the IR brane, the singlets of $SO(4)\times SO(4)$ in the $T_L$ multiplet can couple to $t_R$. Formally, these couplings can be written in terms of the spurions that break the $SO(5)$s into $SO(4)$s, denoted as $S_{u,d} = (0,0,0,0,1)$. The IR Lagrangian is given by:
\begin{equation}\label{eq:IRR}
L_{IR} = \mu \bar{\chi}_L^{u} \Sigma^u S_u  t_R + \mu \bar{\chi}^{d}_L \Sigma^d S_d t_R + h.c. 
\end{equation}
with the projections of the $pNGB$ matrices written explicitly as
\begin{eqnarray} 
\Sigma^{u,d}S_{u,d}=\frac{\sin\left(\frac{h^{u,d}}{f}\right)}{h^{u,d}}\left[h^{u,d}_1,h^{u,d}_2,h^{u,d}_3,h^{u,d}_4,h^{u,d}\cot\left(\frac{h^{u,d}}{f}\right)\right]^T~,~h^{u,d}=\sqrt{h^{u,d}_ah^{u,d}_a}.\label{eq:higgspNGBs}
\end{eqnarray}
We have also assumed that the IR masses respect the $u \leftrightarrow d$ symmetry (similarly to the bulk masses). 

\subsection{The top contribution to the Higgs potential  }

The 4D effective Lagrangian for the top zero mode can be obtained from Eq.~(\ref{eq:IRR}) by inserting the bulk profile of Eq.~(\ref{eq:bulkk}) evaluated on the IR brane. The profile on the IR brane is dominated by the combination $T^0_L =\frac{1}{\sqrt{2}}\left( T^u_L + i T_L^d\right)$ of the two bulk multiplets, and so on the IR brane
\begin{equation}
\chi^d_L \simeq i \chi^0_L~, ~\chi^u_L \simeq \chi^0_L
\end{equation}
The 4D Lagrangian for the top is then:
\begin{eqnarray} 
\mathcal{L}_{\text{top}}=\frac{y_t\, f}{\sqrt{2}} \bar{\chi}^0_L \left(\Sigma^{u}S_u +i \Sigma^{d} S_d \right) t_R 
\end{eqnarray}
where 
\begin{eqnarray} 
\bar{\chi}^0_L=\frac{1}{\sqrt{2}}\left(-i\bar{b}_L,\bar{b}_L,i\bar{t}_L,\bar{t}_L,0\right)\, .
\end{eqnarray}
Substituting these definitions and expanding to second order in the Higgs doublets $h_{1,2}$, where
\begin{eqnarray} 
h_{1,2}=\frac{1}{2}\left(
\begin{array}{c}
\, \, \,ih^u_3+h^u_4\pm i\left(~~i h^d_3+h^d_4\right)\\
-ih^u_1+h^u_2\pm i\left(-i h^d_1+h^d_2\right)
\end{array}
\right)\, ,
\end{eqnarray}
we get the effective Yukawa coupling of the top:
\begin{eqnarray} 
\mathcal{L}_{\text{eff}}=y_t\, h_1~\bar{t}_L t_R.
\end{eqnarray}
We see that the top couples only to $h_1$, so that it contributes a $- \frac{3 y_t^2 M^2_{KK}}{16 \pi^2} h_1h^\dagger_1$ to the 2HDM potential, generating a negative quadratic term in the $h_1$ direction.

\subsection{Lifting the flat direction\label{sec:lift}}

To lift the flat direction we must have a sizable positive mass term of $h_2$. To do that, we break down the contribution of the top partners to the Higgs potential. The top partners populate the $T_L$ multiplet, which is the $({\mathbf 5},\mathbf{1})+(\mathbf{1},{\mathbf 5})$ of $SO(5)\times SO(5)$. As before, we separate $T_L$ into the linear combinations:
\begin{equation}
T_L^{0,1} = \frac{1}{\sqrt{2}}\left(T_L^u \pm i T_L^d\right)~,~
\end{equation}
These multiplets couple to $t_R$ through the pNGB matrices:
\begin{equation} 
\mathcal{L}_{\text{top+topKK}}=\frac{y_t\, f}{\sqrt{2}} \bar{\chi}^{0,1} \left(\Sigma^{u}S_u \pm i \Sigma^{d} S_d \right) t_R + h.c.   \label{eq:KKHiggs}
\end{equation}
where $\chi^{0,1}$ are the left handed Weyl fermions in the Dirac fermions $T^{0,1}_L$. 

The total contribution to the Higgs potential from complete multiplets $T^0_L$ and $T^1_L$ has to be zero, because their linear combinations $T^u_L,\, T^d_L$ form complete $SO(5)\times SO(5)$ multiplets. Furthermore,  even the individual contributions of complete multiplets $T^0_L$ and $T^1_L$ are separately zero. Indeed, the Coleman Weinberg potential depends on $T^{0,1}_L$ through
\begin{equation}
\left| \frac{y_t\, f}{\sqrt{2}}\left(\Sigma^{u}S_u \pm  i \Sigma^{d} S_d \right) \right|^2 =y^2_t f^2\, ,
\end{equation}
which is independent of $h_{1,2}$. Since our zero modes live mostly in $T^0_L$ and all the other modes are at $M_{KK}$, we can now ignore the complete multiplet $T^1_L$ and focus only on the $T^0_L$ contribution, which is cut at $M_{KK}$.

The top partners in $T^0_L$ are:
\begin{eqnarray} 
T^{0}_L=\frac{1}{\sqrt{2}}\left(0,0,-i\bar{t}^{0}_d,\bar{t}^{0}_d, t^{0}_s \right)\, .
\end{eqnarray}
Here $t^{0}_d$ is part of an electroweak doublet and $t^{0}_s$ is an electroweak singlet. Their couplings to the Higgs in Eq.~(\ref{eq:KKHiggs}) can be explicitly written as:
\begin{equation}\label{eq:KKHiggs_full}
\mathcal{L}_{\text{top partners}}=y_t \bar{t}^0_{dL} h^\dagger_2  t_R  +y_t \bar{t}^0_{sL} \left(f-\frac{h^2_1}{2 f} -\frac{h^2_2}{2 f}\right)  t_R +M_{d} \bar{t}^0_{dL}t^0_{dR}+M_{s} \bar{t}^0_{sL}t^0_{sR}\, ,
\end{equation}
while the Higgs potential contribution of this sector is
\begin{equation}
V(h_1,h_2)=  - \frac{3 y_t^2 M^2_{s}}{16 \pi^2} h_1h^\dagger_1+ \frac{3 y_t^2 \left(M^2_{d}-M^2_{s} \right)}{16 \pi^2} h_2h^\dagger_2\, .
\end{equation}

\begin{figure}
\vspace*{-0.5cm}
\begin{center}
\includegraphics[scale=0.3]{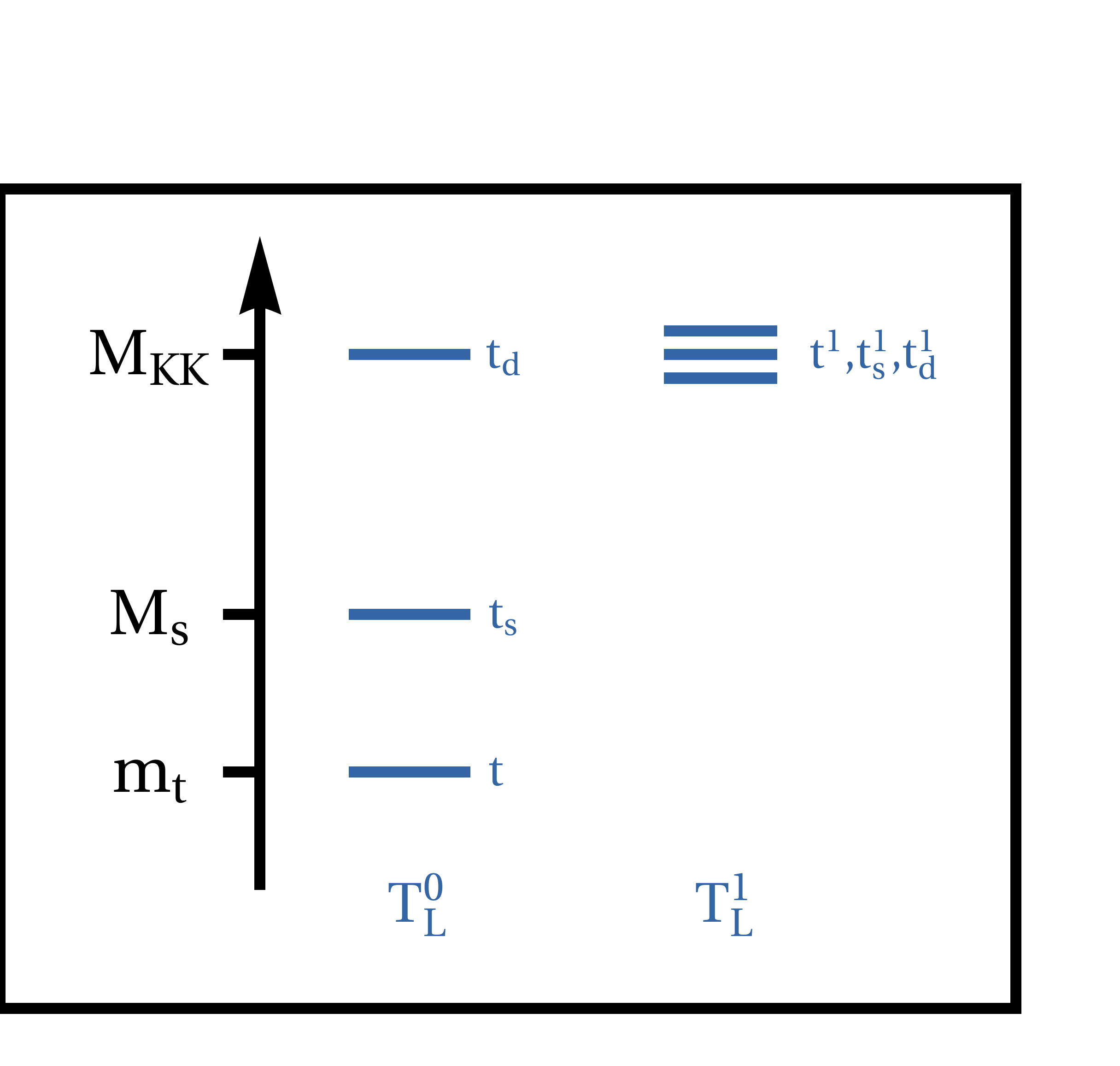}
\end{center}
\vspace*{-0.2cm}
\caption{The top partner spectrum in our model. \label{fig:spec}}
\end{figure}

The top partner spectrum of our model is shown in Fig.~\ref{fig:spec}. We see that $M_s$ acts effectively as a cutoff for the $h_1$ mass term, while both the doublet and the singlet contribute to $h_2$ with opposite signs. For $M_{d}>M_{s}$, we get a positive mass for $h_2$, lifting the flat direction as required. A simple way to achieve this is to take $t^0_{sL}$ as a zero mode as well, and marry it in the bulk and on the IR to a similar right handed fermion in a new multiplet. The b.c. for the top multiplet are thus taken as:
\begin{eqnarray}
(t_L,b_L)&:&  \tilde{\chi}^u_{L}:(-,+) ~,~\tilde{\chi}^d_{L}:(+,+)\, \nonumber\\
(t_{dL},b_{dL})&:&  \tilde{\chi}^u_{L}:(-,+) ~,~\tilde{\chi}^d_{L}:(-,+)\, \nonumber\\
s_L&:&  \tilde{\chi}^u_{L}:(-,+) ~,~\tilde{\chi}^d_{L}:(+,+),
\end{eqnarray}
where the b.c. are given for the left-handed Weyl fermion and the b.c. are flipped for the opposite chirality. The new multiplet $S_R$ is also in a $({\mathbf 5},\mathbf{1})+(\mathbf{1},{\mathbf 5})$ of $SO(5)\times SO(5)$. Its boundary conditions are different from those of $T_L$, as only the singlets of $SO(4)\times SO(4)$ have BC similar to the top
\begin{equation}
\tilde{\chi}^u_{S}:(-,+) ~,~\tilde{\chi}^d_{S}:(+,+)\, ,
\end{equation} 
while the rest have $(-,+)$ BC for both up and down components. This means that the zero mode, which we denote $s^0_R$, exactly mirrors $t^0_{sL}$ and can marry it in the bulk and on the IR.  The bulk mass of the multiplet is chosen such that zero mode is localized sufficiently far from the UV so that $M_{s}\lsim M_{kk}$. We also choose the mixing of the two multiplets to obey the full $SO(5)\times SO(5)$ symmetry so that a new contribution to the Higgs potential is not generated.  The 4D Lagrangian is then exactly Eq.~(\ref{eq:KKHiggs_full}) with $M_{s}$ being a free parameter. With our choice of $M_{s} < M_{d} \sim M_{KK}$, the total radiative Higgs potential is
\begin{equation}\label{eq:cw}
V(h_1,h_2)=  - \frac{3 y_t^2 M^2_{s}}{16 \pi^2} h_1h^\dagger_1+ \frac{3 y_t^2 M^2_{KK}}{16 \pi^2} h_2h^\dagger_2\, .
\end{equation}
This potential lifts the $h_2$ mass to the KK scale, so that:
\begin{equation}
m_{h_2} =  \frac{M_{KK}}{M_s} m_{h1} \approx \frac{M_{KK}}{\sqrt{2}M_s} m_{h}
\end{equation}
where $m_h$ is the measured mass of the SM Higgs (which corresponds to $h_1$ in our scenario).

\section{The 2HDM Potential\label{sec:2HDM}}

In this section we present the Higgs potential in our model. As we will see, this is a 2HDM potential in the decoupling limit\footnote{For another example of a CH model with a 2HDM scalar sector, see~\cite{Mrazek:2011iu}.}. The $h_1$ doublet will serve as the SM Higgs, coupling with approximately SM magnitudes to the W and Z gauge bosons and to the top and bottom sector, while $h_2$ will be heavy. 

\subsection{Mass terms and quartic}
The SM-like $h_1$ gets a negative quadratic term via top loops
\begin{equation}
m^2_{h_1} \sim -\frac{3 y_t^2 M^2_{s}}{16 \pi^2}\, ,
\end{equation}
while the positive quadratic term of the second Higgs is
\begin{equation}
m_{h_2} \sim \frac{M_{KK}}{M_s} m_{h_1}\, ,
\end{equation}
as explained above. 
Additionally, there is a small $h_1h_2$ mixing from the subdominant mode with bulk mass $c-\Delta c$, which we have neglected previously. Including this mode, we get a fermion Lagrangian of the form of Eq.~(\ref{eq:KKHiggs_full}) with the following corrections: 
\begin{equation}
h_1\rightarrow h_1+\epsilon h_2~,~h_2\rightarrow h_2+\epsilon h_1\, ,
\end{equation}
where $\epsilon\sim {\left(\frac{R}{R'}\right)}^{2\Delta c}\sim\frac{1}{40}$ is the mixing angle of the subdominant combination in the top zero mode. The resulting correction to the 2HDM potential is subdominant except for a possibly important mixing term $-m^2_{12}\, h_1h_2$ which was neglected in Eq.~(\ref{eq:cw}). We estimate this term to be
\begin{equation}
m^2_{12} \sim \epsilon m^2_{h_2}\, .
\end{equation}

The only tree-level part of the 2HDM potential is the quartic term
\begin{equation}
 \lambda \left(h_1^\dagger h_1 - h_2^\dagger h_2\right)^2 +\lambda \text{ tr}\left(h_1h_1^\dagger - h_2h_2^\dagger\right)^2\, ,
\end{equation}
with $\lambda$ given by  Eq.~(\ref{Eq:5Dquartic}). 

\subsection{Matching to the general 2HDM potential}

The general 2HDM Lagrangian \cite{2HDM} is
\begin{eqnarray}
\mathcal{L}_{h_{12}}&=&m^2_{h_1}\,h^\dagger_1 h_1+m^2_{h_2}\,h^\dagger_2 h_2-m^2_{12} \left(h^\dagger_1 h_2 + \text{h.c.} \right)+ \nonumber \\
&&\,+\, \frac{1}{2}\lambda_1\,{\left(h^\dagger_1 h_1\right)}^2+\frac{1}{2}\lambda_2\,{\left(h^\dagger_2 h_2\right)}^2+\lambda_3\,{\left(h^\dagger_1 h_1\right)}{\left(h^\dagger_2 h_2\right)}+\lambda_4\,{\left(h^\dagger_1 h_2\right)}{\left(h^\dagger_2 h_1\right)} \nonumber\\
&&+ ~~\text{other CP violating quartics}\, .
\end{eqnarray}
Matching this to our case, we have $\lambda_1=\lambda_2=-2\lambda_3=-2\lambda_4=4\lambda$, while the other CP violating quartics are zero - exactly the same as in the Higgs sector of the MSSM.
Since $|m^2_{12}|<|m^2_{h_1} |<|m^2_{h_2} |$, we are in the decoupling limit with small $\tan \beta$. Indeed, minimizing the potential we get:
\begin{eqnarray}
\tan \beta \sim \,\epsilon \sim \frac{1}{40}\, ,
\end{eqnarray}
so that $m^2_A=m^2_{H^{\pm}}=m^2_{h_2}$.
In the decoupling limit, we can just fit the quartic to its experimental value
\begin{equation}
\frac{1}{2}\, g^2_* \theta^2_6=2\lambda\approx 0.13\, .
\end{equation}
For $g_*\gsim 3$ we get that the bulk $SO(5)\times SO(5)$ breaking parameter is $\theta_6 \sim 0.1$. 
The physical Higgs masses in the decoupling limit are just:
\begin{equation}
m_h \simeq m_{h_1}~,~m_H \simeq m_{h_2}\, ,
\end{equation}
while the alignment angle is
\begin{equation}
\cos \left(\alpha-\beta\right) = \tan \beta \, \frac{m_h}{\sqrt{m^2_H-m^2_h}}\, .
\end{equation}
Since our $\cos \left(\alpha-\beta\right)$ is small, our decoupling limit leads to alignment: the vector bosons couple to $h$ with SM-like magnitudes, while their coupling to $H$ is supressed. In other words, $h$ is almost completely SM-like, while $H$ is both heavy and inert up to $\mathcal{O}(\frac{1}{40})$.

 \section{Phenomenological Consequences\label{sec:pheno}} 

The BSM phenomenology of our model can be largely divided into standard composite Higgs phenomenology, which implies all the generic features of composite Higgs models, and additional features which are specific to our model.  The standard composite Higgs phenomenology has been covered extensively in the literature (see \cite{Bellazzini:2014yua}), and here we only briefly summarize the main predictions:
\begin{itemize}
\item Fermionic top partners: New vector-like quarks that would appear in direct searches. In our model there is a light top partner - $t^0_s$ with SM quantum numbers $({\bf 3},{\bf 1}, 2/3)$, which is lighter than the rest of the KK states. This state cancels the top loop and so its mass is directly connected to the tuning. Current constraints are $M_{s} \gsim 1.1$ TeV \cite{Aaboud:2017zfn,Sirunyan:2017pks}, which corresponds to a tuning of ${\cal O }(20\%)$. In contrast with standard CH models where Higgs and EW constraint favor a heavier top partner $M_T \gsim 1.5-2\text{ TeV}$, the top partners in our model can be light. We assume the rest of the top KK modes are much heavier: $M_{KK} > 3$ TeV.
\item Gauge partners: Heavy new EW vectors that would appear as resonances at the LHC. The gauge partners also directly contribute to the tuning, and their mass has to be $M_{V} \gsim 2.5$ TeV. In this model we assume that the necessary ${\cal O }(20\%)$ tuning in the Higgs sector is obtained via the cancellation of the top and gauge loops. 
\item Higgs data: Due to the pNGB nature of the Higgs, the coupling of the Higgs differ from the SM expectation by ${\cal O}\left(\frac{v^2}{f^2}\right)$. This is also the main deviation in the Higgs signals in our model.
\item EW precision data: In composite Higgs models with custodial protection, the dominant effect is on the S parameter due to tree-level mixing with the gauge partners. At the loop level, the change in the Higgs coupling and the new top partners also have a measurable effect. This is true for our model as well.
\item Flavor: The new composite states can mediate FCNC. These are suppressed if the SM fermions are partially composite, but the suppression is usually not sufficient for the composite Higgs to be the solution of the hierarchy problem. To avoid these constraints, CH models with various flavor symmetries have been considered \cite{FlavorSymmetries}. Addressing these constraints is beyond the scope of this paper, but it is plausible that similar symmetries can be implemented in our model for the same purpose. 
\end{itemize}

The main non-standard features of our model are the doubled KK spectrum and the extended Higgs sector.

\underline{Double KK spectrum.} All the KK modes in our model appear in both the up and down sites. The splitting between two such modes is ${\cal O}(\theta_6 M_{KK})\sim 300$ GeV for fermion KK modes and ${\cal O}(\theta^2_6 M_{KK})\sim 30$ GeV for vector KK modes. To obtain the measured quartic we require $\theta_6 \sim 0.1$. The small splitting between the modes is a clear prediction of our model. 

\underline{Extended Higgs sector.} The Higgs sector of our model is a 2HDM where the second doublet is almost inert. The mass of the second doublet depends on the ratio between the light top partner and the KK scale:
\begin{equation}
m_{h2} \sim \frac{M_{KK}}{\sqrt{2}M_S} \, m_h
\end{equation}
For reasonable choice of parameters, we get $m_{h2} \sim 300-500$ GeV.  The phenomenology of the states in this doublet depends on the embedding of the quarks in our model. If we assume that all the quarks couple to $h_1$ similarly to the top, then the coupling to the second doublet are suppressed by the mixing which is $\mathcal{O}(\frac{1}{40})$. This suppresses all single production processes for the scalar states in the second doublet. The charged Higgs can additionally be produced in pairs via an unsuppressed Drell-Yan process. Further analysis is required to obtain precise bounds and discovery potential for these states.

\section{Conclusions\label{sec:Conclusions}}

Composite Higgs models are among the most exciting viable extensions of the SM. One of the long-standing issues with such models is the tuning necessary to obtain a realistic Higgs potential with a 125 GeV Higgs mass. In this paper we presented a novel class of models where this tuning can be reduced in the presence of a tree-level quartic for the Higgs. The model originates from a 6D orbifold producing two pNGB Higgs doublets which naturally obtain a collective quartic due to the 6D gauge interactions. The simplest and most useful implementation is based on its 5D two-site deconstructed version, where the bulk gauge group for the minimal model is $SO(5)\times SO(5)$, with a bulk link field breaking the group to the diagonal, in addition to boundary breakings ensuring the presence of the two light Higgs doublets. This is the first example of a holographic composite Higgs model with a tree-level quartic for the Higgs. Importantly, the magnitude of the quartic can be adjusted by dialing the parameters of the theory which can result in the SM quartic dominantly arising from the tree-level contribution. This in turn allows us to suppress all loop effects, leading to the reduction of the tuning. The main experimental consequence of such models (beyond the standard signals of composite Higgs models such as colored top partners, gauge partners, etc) is the doubling of the KK spectrum. The splitting among the KK modes is related to the same parameters setting the magnitude of the quartic. We find that for a realistic model with reduced tuning, the splitting among the doubled KK states has to be relatively small (of order tens of GeV for the vector KK modes and hundreds of GeV for the fermionic KK modes). In addition, the Higgs sector has to contain two Higgs doublets, which must be in the decoupling limit in order to correctly reproduce the measured Higgs phenomenology. Finally, unlike the CH case, direct searches for fermionic top partners provide the most stringent bounds on naturalness in our model, and so we expect them to lie just around the corner within the LHC reach.

\section*{Acknowledgements}

We are grateful to Nima Arkani-Hamed for collaboration at the early stages of this project. We also thank Kaustubh Agashe, Brando Bellazzini, Zackaria Chacko, Jack Collins, Yuri Shirman, Raman Sundrum and John Terning for useful discussions. We thank the Mainz Institute for Theoretical Physics for its hospitality while this work was in progress. C.C. and M.G. also thank the Aspen Center for Physics - supported in part by NSF-PHY-1607611 - for its hospitality while working on this project. C.C. and O.T. are supported in part by the NSF grant PHY-1719877. MG is supported by the NSF grant PHY-1620074 and the Maryland Center for Fundamental Physics.

\end{document}